\documentclass[aps,prb,reprint,superscriptaddress,showpacs]{revtex4-1}
\usepackage{hyperref}
\usepackage{graphicx}
\usepackage{epstopdf}
\usepackage{amssymb}
\usepackage{amsmath}
\usepackage{bbm}
\usepackage{color}
\usepackage{xfrac}
\usepackage{datetime}

\newlength{\tlength}

\newcommand\solidrule[1][0.5cm]{\rule[0.5ex]{#1}{.8pt}}
\newcommand\dashedrule{\mbox{%
  \solidrule[1mm]\hspace{1mm}\solidrule[1mm]\hspace{1mm}\solidrule[1mm]}}

\begin{document}

\title{Closing the gap between spatial and spin dynamics of electrons at the metal-to-insulator transition}

\author{J. G. Lonnemann}
\author{E.P. Rugeramigabo}
\author{M. Oestreich}
\email{oest@nano.uni-hannover.de}
\author{J. H\"ubner}
\email{jhuebner@nano.uni-hannover.de}
\affiliation{Institute for Solid State Physics, Leibniz Universit\"at Hannover, Appelstr. 2, 30167 Hannover, Germany}
\date{\today}

\begin{abstract}
We combine extensive precision measurements of the optically detected spin dynamics and magneto-transport measurements in a contiguous set of n-doped bulk GaAs structures in order to unambiguously unravel the intriguing but complex contributions to the spin relaxation at the metal-to-insulator transition (MIT). Just below the MIT, the interplay between hopping induced loss of spin coherence and hyperfine interaction yields a maximum spin lifetime exceeding 800~ns. At slightly higher doping concentrations, however, the spin relaxation deviates from the expected Dyakonov-Perel mechanism which is consistently explained by a reduction of the effective motional narrowing with increasing doping concentration. The reduction is attributed to the change of the dominant momentum scattering mechanism in the metallic impurity band where scattering by local conductivity domain boundaries due to the intrinsic random distribution of donors becomes significant. Here, we fully identify and model all intricate contributions of the relevant microscopic scattering mechanisms which allows the complete quantitative modeling of the electron spin relaxation in the entire regime from weakly interacting up to fully delocalized electrons.

\end{abstract}


\maketitle

\section{Introduction}

Spin based semiconductor electronics is a long lasting vision which drives basic research in many complementary disciplines of physics.\cite{Wolf.Science.2001} One of the open fundamental questions regards the exact physical mechanisms evoking spin relaxation at the metal-to-insulator transition where experiments in GaAs reveal extremely long spin relaxation times $\tau_{\rm s}$ at low temperatures. Electron spin relaxation in GaAs is for high doping densities and/or high temperatures clearly effectuated by the prominent Dyakonov-Perel (DP) mechanism while hyperfine interaction dominates at very low doping densities and low temperatures. Both situations are perfectly understood by theory and confirmed by experiments. In contrast, the theory of electron spin relaxation at the MIT is still discussed very controversially. Shklovskii predicted 
that the traditional DP mechanism is not only valid at high doping densities but also for metallic samples close to the MIT.\cite{Shk2006} Tight binding calculations without adjustable parameters predicted one year later a spin relaxation time exceeding 1$\mu$s at the MIT with the Elliot-Yafet mechanism as the major source for the loss of spin coherence.\cite{Tam2007} Models taking into account hopping together with effective spin-orbit interaction for the impurity system in n-doped GaAs predicted a square-root dependency of $\tau_{\rm s}$ with doping density and matched the existing experimental data by adjusting the strength of the cubic Dresselhaus spin splitting parameter $\gamma_{\rm D}$.\cite{Int2012, Wellens.PRB.2016}
However, the available experimental data on $\tau_{\rm s}$ strongly scatters around the MIT and cumulative experimental data even suggest an abrupt drop of $\tau_{\rm s}$ at the insulating side of the MIT \cite{Dzh2002} which is not predicted by theory.

\begin{figure}[t]
    \includegraphics[width=\linewidth]{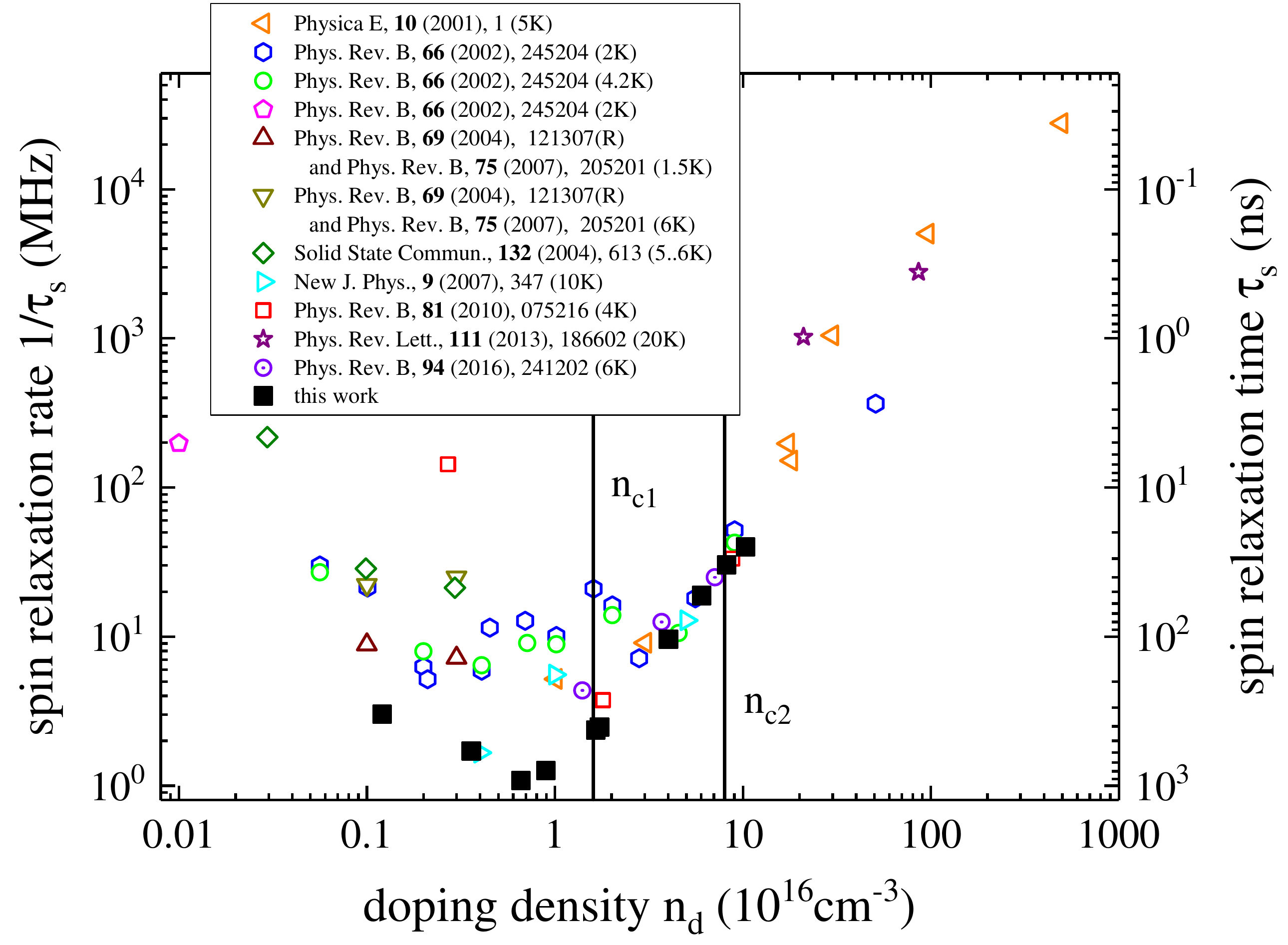}
        \caption{\label{fig:DopLit} (color online) High precision, low temperature measurements ({\tiny $\blacksquare$}) of the electron spin relaxation time $\tau_{\rm s}$ in n-GaAs as function of donor density compared to literature values of $\tau_{\rm s}$ in the limit of negligible magnetic fields.\cite{Awschalom.PhysicaE.2001,Dzh2002, Colton.PRB.2004, Colton.PRB.2007, Colton.SSC.2004, Furis.NJP.2007, Roemer.PRB.2010,Berski.PRL.2013, Belykh.PRB.2016} \vspace{\tlength}}
\end{figure}
Electron spin coherence in n-doped GaAs is inextricably linked to the dynamics in real and momentum space via the local hyperfine fields (HF) $\mathbf{\Omega} (\mathbf{r})/\mu_{\mathrm{B}}$, the momentum depended effective spin splitting $\hbar\mathbf{\Omega} (\mathbf{k})$, and the corresponding correlation time $\tau_{\rm c}$.\cite{Pines.PR.1955} The latter is mainly given by the type and the duration of interaction which result for example from variable range hopping (VRH) for predominantly localized carriers or from the effective momentum scattering times $\tau_{\rm p}$ in the case of free carriers. Understanding spin relaxation at the MIT is therefore directly linked to a fundamental understanding of the spatial dynamics of the electrons at the MIT.
Here, we present optical high-precision measurements of $\tau_{\rm s}$ together with extensive temperature and magnetic field dependent transport measurements which reveal the relevant correlation times and momentum scattering mechanisms. All measurements are performed on n-doped, $2\:\mathrm{\mu m}$ thick, high quality GaAs:Si epilayers grown by molecular beam epitaxy with nominal doping densities ranging from $n_{\rm d}=10^{15}\mathrm{cm^{-3}}$ to $10^{17}\mathrm{cm^{-3}}$, enclosed by adapted n-doped top and bottom layers in order to reduce surface and interface effects. The doping regime covers the complete dynamics from strongly localized carriers up to the fully degenerate case and allows a coherent \emph{quantitative} modeling of all contributing major mechanisms.

The spin relaxation times are carefully measured by Hanle depolarization of the photoluminescence at 6.5\,K. The black squares in Fig.~\ref{fig:DopLit} depict the measurements of $\tau_s$ versus doping density in comparison to experimental values from literature. The two vertical lines denote the critical densities $n_{\rm c1}=1.6\times 10^{16}\,\mathrm{cm}^{-3}$ and $n_{\rm c2}=8\times 10^{16}\,\mathrm{cm}^{-3}$, i.e., the point of finite conductivity at the Mott MIT in the limit of zero temperature ($n_{\rm c1}$) and the onset of the impurity band hybridizing with the conduction band ($n_{\rm c2}$), respectively \cite{Ben1987, Romero.PRB.1990}. First of all, our measurements show that there is no sharp change of $\tau_s$ at the MIT but the change is gradual. Secondly, the measurements on our samples yield in comparison to the data from literature an upper bound for $\tau_s$ which is most apparent slightly below the MIT where data from literature scatters by nearly two orders of magnitude.
\section{Electron dynamics}

\begin{figure}[t]
    \includegraphics[width=\linewidth]{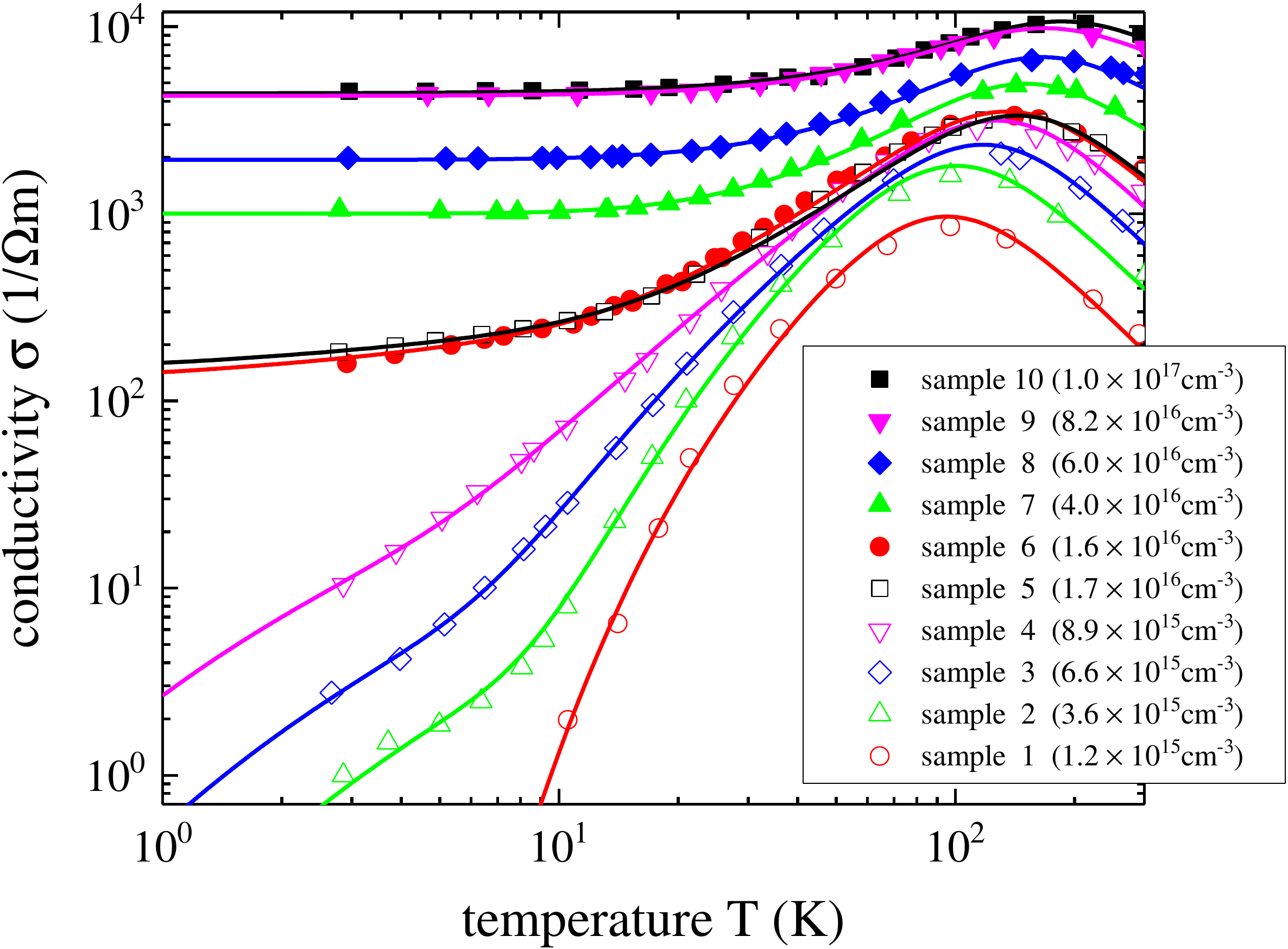}
        \caption{\label{fig:cond} (color online) Temperature dependence of the conductivity $\sigma$. The solid lines are fits according to Eq.~\ref{eq:sigma} with the respective contributions depending on the doping density.}\vspace{\tlength}
\end{figure}
\begin{figure}[t]
    \includegraphics[width=\linewidth]{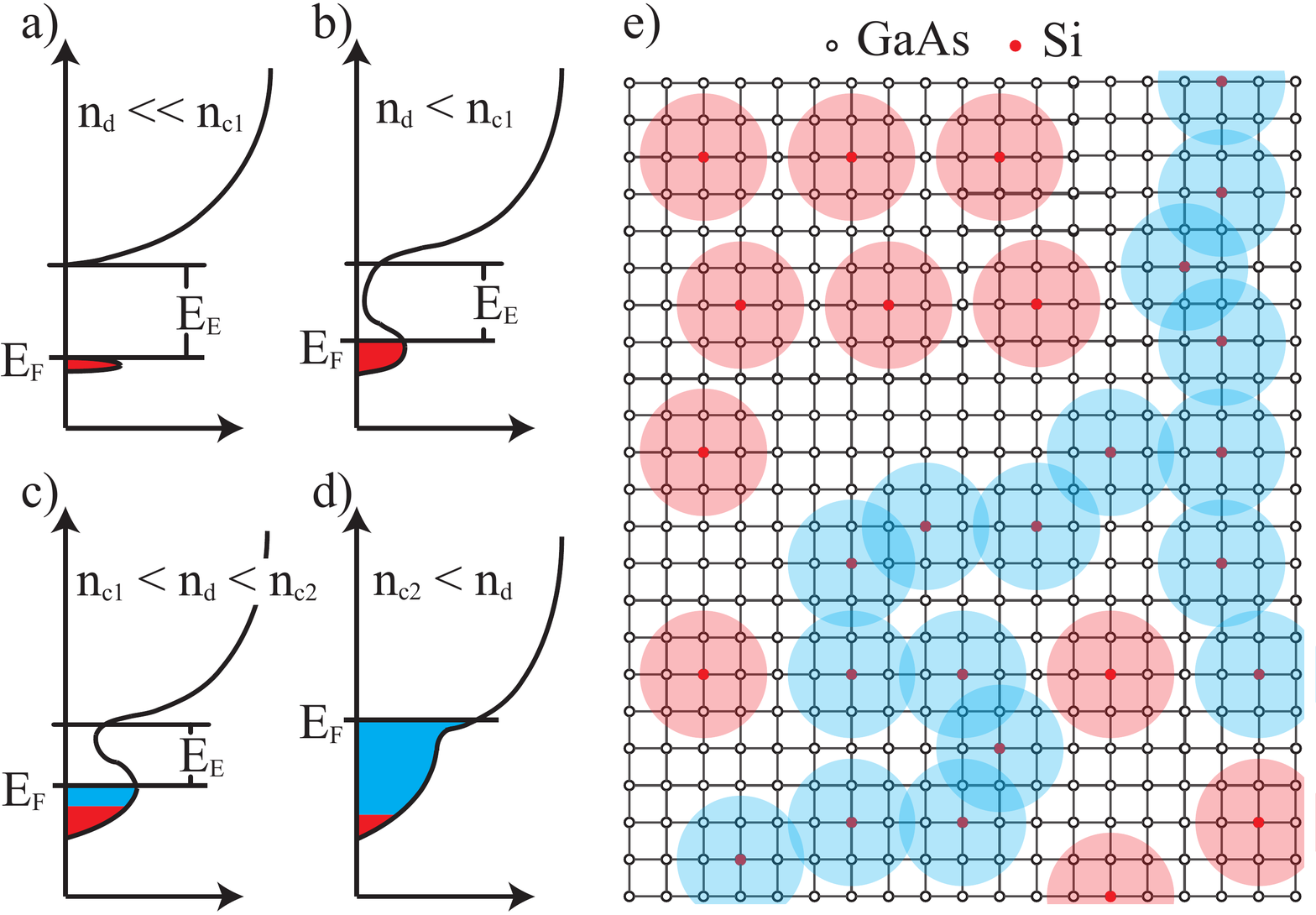}
        \caption{\label{fig:model}(color online) (a-d) Effective density of states at different doping concentrations. The density of localized states is depicted in red, delocalized states are shown in blue. e) Schematic picture of the coexistence of variable range hopping and filamentary electron transport along percolation paths at the MIT.}\vspace{\tlength}
\end{figure}
Next, we study on the \emph{same} samples the temperature dependence of the conductivity $\sigma$ (see Fig.~\ref{fig:cond}) and
analyze the data quantitatively in terms of a two channel transport model \cite{Wal1982}
\begin{equation}
\sigma(T)=\sigma_{\rm di}(T) + \sigma_{\rm cb}(T)
\label{eq:sigma}
\end{equation}
with contributions from a below band gap, doping induced conductivity channel $(\rm di)$ and the non-degenerate parabolic conduction band $(\mathrm{cb})$. The measurement of the temperature dependence is necessary in order to characterize the transport in the impurity band and to distinguish between the influence of the doping induced conductivity channel and the conduction band. Figure~\ref{fig:model} depicts schematically the underlying model: At low doping densities ($n_{\rm d} \ll n_{\rm c1}$), (a) the donors are well separated and mainly localized while with increasing doping density ($n_{\rm d} < n_{\rm c1}$), (b) the overlap of the individual donors increases, the density of impurity states broadens, and the Fermi energy $E_{\rm F}$ of the weakly interacting electrons increases. This regime is dominated by variable range hopping where the impurity and band tail states are separated from the conduction band by an effective doping dependent excitation energy $E_{\rm E}$. For $n_{\rm c1}\le n_{\rm d} \le n_{\rm c2}$, (c) the density of impurity states broadens and metallic filamentary electron transport becomes possible. For $n_{\rm d} \ge n_{c2}$, (d) the Fermi energy lies in the conduction band and the electron transport remains metallic.

For elevated temperatures, $\sigma(T)$ is dominated for all doping densities by the conduction band conductivity $\sigma_{\rm cb}(T)$ which first increases with temperature since ionized impurity scattering decreases and/or electrons are thermally excited from the impurity and band tail states to the conduction band. At even higher temperatures $\sigma(T)$ decreases again due to the impact of polar optical phonon scattering \cite{Blakemore.JAP.1982}. The conductivity in the conduction band can be easily calculated by
\begin{equation}
\sigma_{\rm cb}(T)=\left[\mu_\mathrm{II}^{-1}(T)+\mu_\mathrm{PO}^{-1}(T)\right]^{-1} n_\mathrm{d} \:\mathrm{e}^{-\frac{E_\mathrm{E}}{k_\mathrm{B} T}},
\label{eq:cond}
\end{equation}
where the mobility for scattering on ionized impurities is given by $\mu_\mathrm{II}\left(T\right)=A_\mathrm{II}^{RT}/n_\mathrm{d} (T/300\mathrm{K})^{3/2}$ and on polar optical phonons by $\mu_\mathrm{PO}\left(T\right)=\mu_\mathrm{P}^{RT} \: (T/300\mathrm{K})^{-2.3}$ where $A_\mathrm{II}^{RT}/n_\mathrm{d}$ and $\mu_\mathrm{P}^{RT}$ are the particular mobilities at room temperature. \cite{Blakemore.JAP.1982}. The efficiency of the Rutherford type ionized impurity scattering drops with increasing thermal velocity of the electrons while for scattering on phonons the number of available scattering centers increases with temperature. Consequently, the importance of both processes with temperature is opposed.

At low temperatures, the situation is quite different. For doping densities $n_{\rm d} < n_{\rm c1}$, transport is dominated by hopping. The probability $P_{ij}$ of an electron hopping from an impurity $i$ to $j$ decreases exponentially with the inter-donor distance $R_{ij}$ due to the reduced overlap of their wave functions with the effective Bohr radius $a_{\rm d} = a_{\rm Bohr}\: E_{\rm Ryd} / (\epsilon_{\rm r} E_{\rm d})$ where $\epsilon_{\rm r}=12.35$ is the background relative dielectric constant \cite{Strzalkowski.APL.1976} and $E_{\rm d} = 5.8$~meV the donor binding energy in GaAs \cite{Almassy.SSC.1981}. The energy mismatch between the two sites $(\epsilon_{ij} = |E_i -E_j| ))$ is constrained to $\epsilon_{ij} \leq k_{\rm B} T$ whereas the probability of finding such a pair of sites increases with $R_{ij}$. These two opposing dependencies result in an optimal hopping distance $R_{\rm opt}  = [9 a_d /( 8\pi \mathrm{N}_\mathrm{E_{\rm F}} k_{\rm B} T ) ]^{1/4}$ and is thus referred to as variable range hopping defining the process specific diffusion constant $D_{\rm hop}=R_{\mathrm{opt}}^2/(6 \tau_{\rm hop})$ \cite{Nan2015}. Here, $\mathrm{N}_\mathrm{E_F}$ is the density of states at the Fermi energy. The diffusion constant $D_{\rm hop}$ is accessible from the transport measurements and will be used later in order to specify the hopping time $\tau_{\rm hop} $. The average distance traveled per hop increases with decreasing temperature due to the narrowing of the allowed energy range leading to
\begin{equation}
\sigma_{\rm hop}(T)=\sigma_\mathrm{0}\left(\mathrm{N}_\mathrm{E_F}\right)T^{-1/2} \: \mathrm{e}^{-\left(T_\mathrm{0}\left(\mathrm{N}_\mathrm{E_F}\right)/T\right)^{1/4}},
\label{eq:condhh}
\end{equation}
with $\mathrm{N}_\mathrm{E_F}$ as the only free parameter which is quantified in the literature implicitly by $T_\mathrm{0}$ and $\sigma_\mathrm{0}$
\footnote{The exact relations are given by $T_0=512 / (9\pi a_0^3 k_B \mathrm{N}_\mathrm{E_F})$ and $\sigma_0 T^{-1/2}=e^2 R_{\rm opt}^2 \nu_\mathrm{H} \mathrm{N}_\mathrm{E_F}/6$ \cite{Nan2015}. The attempt rate is given by the phonon frequency $\nu_\mathrm{H}=8.8\:\mathrm{THz}$ \cite{Pat1984}.} and compares well with values obtained for comparable samples \cite{Emelanenko.SPS.1973, Ben1987}.
For all samples below the MIT (samples S1 to S4), the conductivity is fully described by Eqs.~\ref{eq:sigma}-\ref{eq:condhh}. Figure~\ref{fig:cond} shows the characteristic temperature activated VRH at low temperatures and the prominent increase of $\sigma$ due to excitation of carriers into the conduction band at elevated temperatures. For sample S1 the temperature induced excitation of carriers into the conduction band is observed as well but the low temperature value of $\sigma$ is so small that $\rm N_{E_F}$ --needed below for the calculation of $\tau_s$-- has to be extrapolated.
\begin{figure}[t]
    \includegraphics[width=\linewidth]{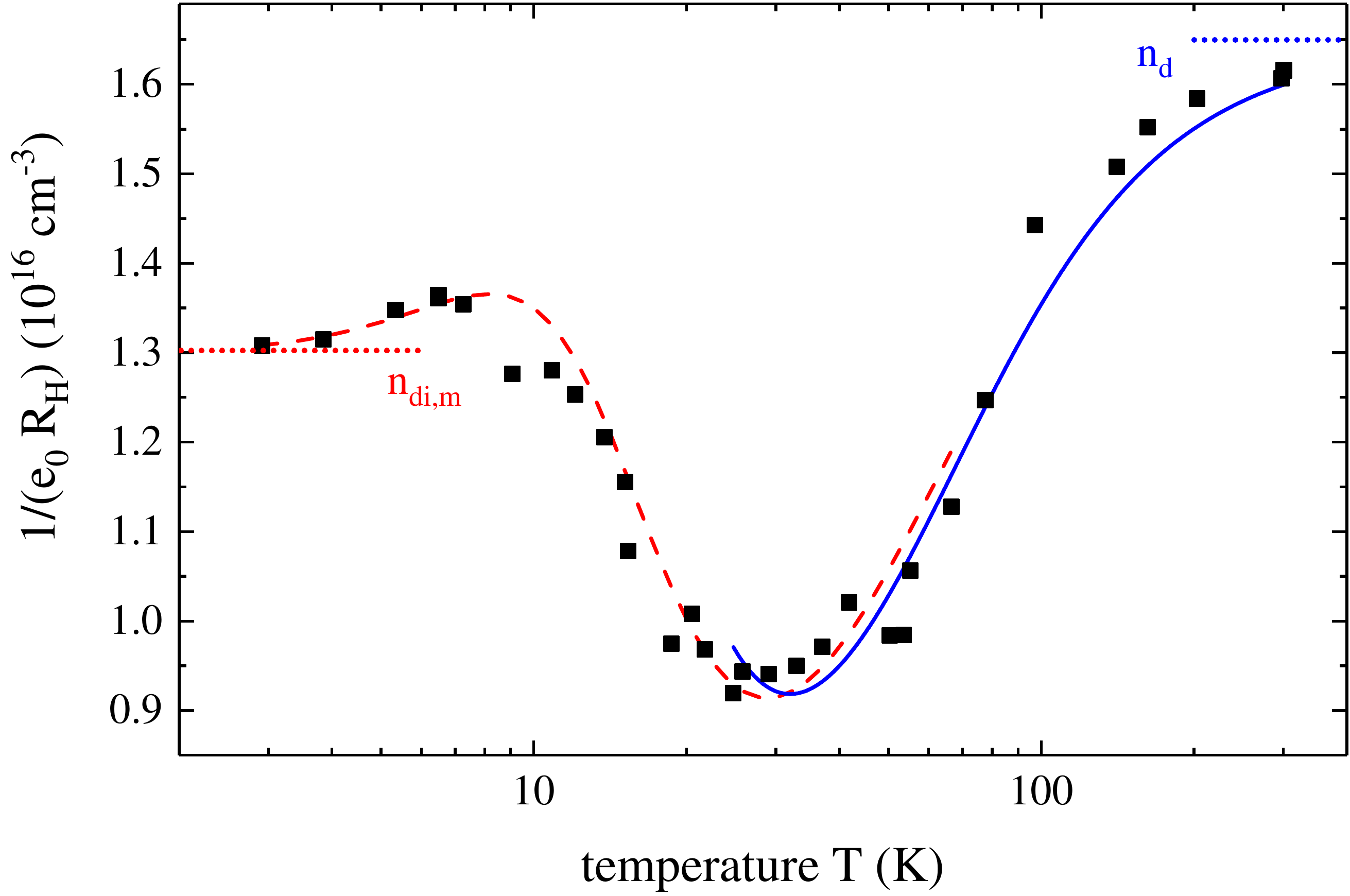}
        \caption{\label{fig:RH} (color online)
        Temperature dependence of the measured inverse Hall resistance exemplary shown for sample S6. The apparent minimum occurs for $n_{\rm di} \mu_{\rm di}=n_{\rm cb}\mu_{\rm cb}$. The two lines are fits according to Eq.~\ref{eq:RH} whereby the high temperature fit (blue solid line) neglects localization and yields $n_{\rm d}$ which in turn enters as a constant to fit the low temperature data (red dashed line). Here, $n_{\rm di,m}$ denotes the carrier concentration related to the non vanishing, zero temperature conductivity $\sigma_{\rm m}^{0}$.}\vspace{\tlength}
\end{figure}

For low temperatures and doping densities slightly above $n_{\rm c1}$, the  conductivity in the impurity channel turns into a mixing of metallic $\sigma_{\rm m}^0$ and hopping conductivity, $\sigma_{\rm di} = \sigma_{\rm m}^0 + \sigma_{\rm hop}$. Figure~\ref{fig:RH} shows exemplary the measured $R_H$ for sample S6 with a doping concentration at the MIT for temperatures between 3 and 300~K. In general $R_\mathrm{H}(T)$ is not directly proportional to the total number of carriers but rather given by \cite{Wal1982}
\begin{equation}
R_\mathrm{H}(T) = \frac{1}{\mathrm{e}_0} \frac{\left(n_{\rm di}(T) + n_{\rm cb}(T) \eta_\mu^2\right)}{\left(n_{\rm di}(T) + n_{\rm cb}(T) \eta_\mu\right)^2},
\label{eq:RH}
\end{equation}
where $\mathrm{e}_0$ is the elementary charge and $\eta_\mu=\mu_{\rm cb}/\mu_{\rm di}$ is the ratio between the conduction band mobility $\mu_{\rm cb}$ and the mobility related to the impurity states $\mu_{\rm di}$. The total electron density is divided into densities related to impurity states $n_{\rm di}$ and the non-degenerate conduction band $n_{\rm cb}$ whereby the thermal activation of carriers from the impurity states to the conduction band is taken into account by \cite{Blakemore.SCStat.1987}
\begin{equation*}\label{eq:activation}
n_{\rm cb}(T)= 2 n_{\rm d} / \left(1+\sqrt{1+4 n_{\rm d} / n_{\rm cb,eff} \cdot e^{E_{\rm d}/(k_{\rm B} T)}}\right),
\end{equation*}
where $n_{\rm cb,eff} = 2 (2 m_e^* k_{\rm B} T)^{3/2} h^{-3}$ is the effective density of states\cite{Blakemore.SCStat.1987}.

Fitting the temperature dependent measurements of $\sigma$ and $R_H$ within the two channel transport model yields all relevant transport parameters which are summarized for completeness in Table~1 for all ten samples. In the following, this quantitative description of the transport correlation times and scattering mechanisms is used to calculate $\tau_s$.

\section{Spin dynamics}

We start with the case of doping densities $n_\mathrm{d} \geq n_\mathrm{c2}$ where the dominating spin relaxation mechanism in bulk III-V semiconductors is of DP type in the motional narrowing regime, i.e., $\tau_{\rm c} \ll \mathbf{\Omega}(\mathbf{k})$. Thus the spin relaxation rate for the degenerate case and low temperatures is given by\cite{Dya1972}
\begin{equation}
 \tau_\mathrm{s}^{-1}=\tfrac{2}{3}\left( \overline{\Omega^2} \; \tau^*\right)_{E_{\rm F}}=\tfrac{96}{35} \; \gamma_{\mathrm{D}}^2 \; {\pi^4 \; n_{\rm d}^2 \; \tau_{\rm c}}/{\hbar^2},
\label{eq:DP1}
\end{equation}
where $E_{\rm F}$ denotes the Fermi energy of an idealized conduction band
and $\tau_{\mathrm{c}}= \tau_\mathrm{p}/\gamma_3$ is the correlation time. The  scattering factor $\gamma_{3}$ relates the transport momentum scattering time $\tau_{p}$ with the correlation --or momentum isotropization-- time $\tau_{c}$ \cite{Dya1972} and depends on the momentum scattering mechanism being unity for isotropic momentum scattering and 6 for small angle scattering in the case of ionized impurities \cite{Pikus.OR.1984}. Sample S10 (and S9) is doped significantly above $n_{\rm c2}$ wherefore at low temperatures ionized impurity scattering dominates \cite{Chattopadhyay.RMP.1981} and the measured spin relaxation rate is nearly temperature independent. Hence, by fitting Eq.~\ref{eq:DP1} with $\gamma_3=6$ to the measured low temperature spin relaxation rates of samples S9 and S10 using the experimentally determined scattering times $\tau_\mathrm{p}$ leaves the Dresselhaus constant $\gamma_{\mathrm{D}}$ as the only free parameter. In the literature, most published experimental \cite{Eng2013,Dre1992,Aro1983} and theoretical \cite{Fu2008,Kna1996,May1993} values of $\gamma_{\mathrm{D}}$ scatter strongly. Our experiment consistently yields with high accuracy $\gamma_{\mathrm{D}} = 19.0(5)\:{\rm eV\,\AA}^3$ which is used for all further calculations in the following and lies well within the previously published values.

\begin{figure}
    \includegraphics[width=\linewidth]{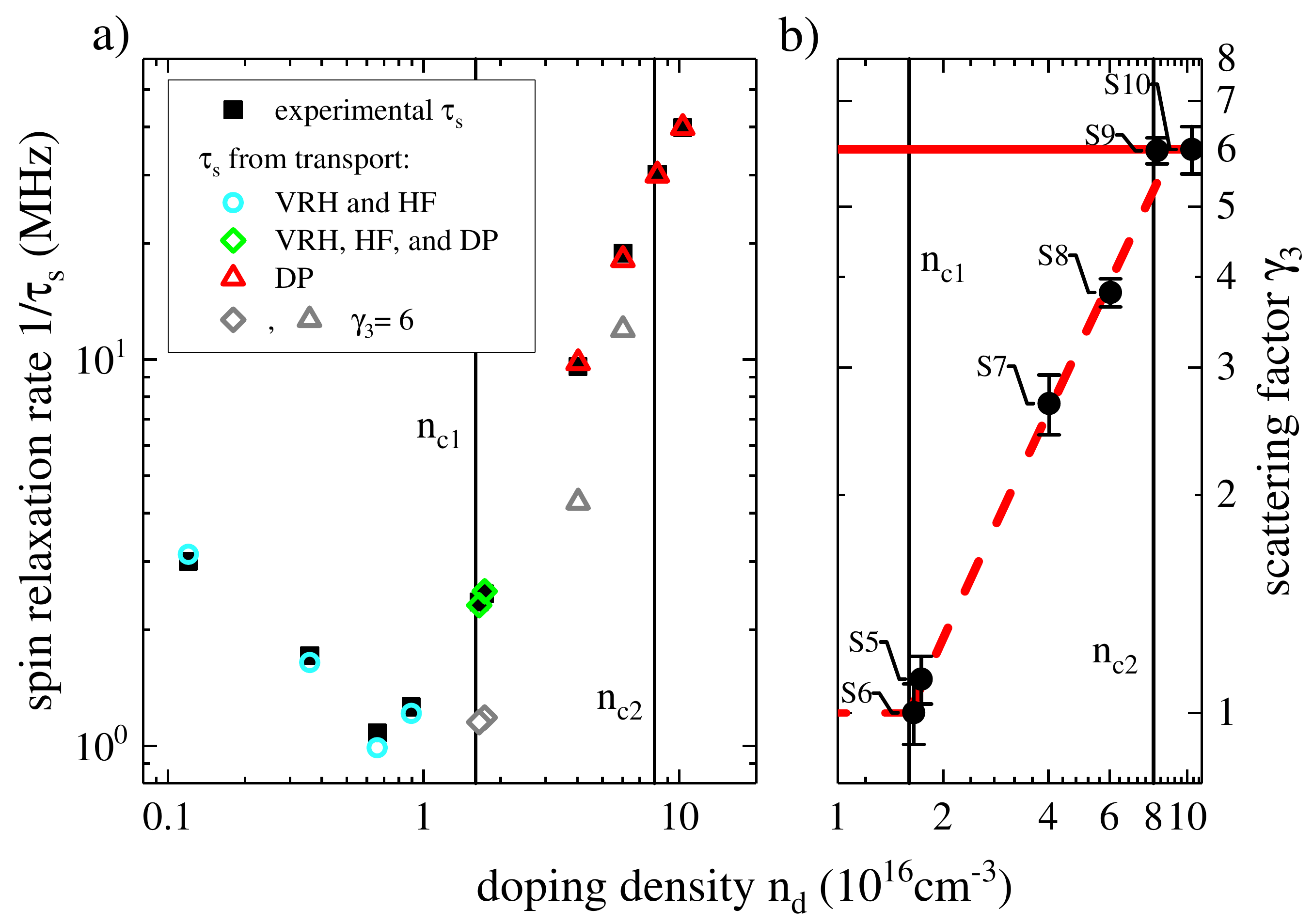}
        \caption{\label{fig:dopdep} (color online) (a) Spin relaxation rate measured at 6.5\,K ($\blacksquare$) in comparison to $\tau_s^{-1}$ calculated from the experimentally measured transport hopping and scattering times. The spin relaxation mechanism changes below the MIT with increasing $n_{\rm d}$ gradually from the HF regime to VRH (\textcolor{cyan}{$\circ$}). Around $n_{\rm c1}$, the spin relaxation results from a combination of HF interaction, VRH, and DP (\textcolor{green}{$\Diamond$}) but already for doping densities between $n_{\rm c1}$ and $n_{\rm c2}$ DP becomes the most dominant spin relaxation mechanism which persist for all higher doping densities (\textcolor{red}{$\triangle$}). The gray symbols denote $\tau_s^{-1}$ calculated with a constant scattering factor $\gamma_3$ which clearly does not reproduce the experimental data. (b) Dependence of $\gamma_3$ on $n_{\rm d}$ that is required to match the measured and the calculated $\tau_s^{-1}$. Between $n_\mathrm{\rm c1}$ and $n_\mathrm{\rm c2}$, a clear linear transition from isotropic, angle independent scattering $\gamma_3 = 1$ towards small angle ionized impurity scattering  $\gamma_3 = 6$ occurs. The linear fit (\textcolor{red}{\dashedrule}) is set to unity at the MIT with $\gamma_3 = 6.67(1)\times10^{-17} (n_{\rm d}-n_{\rm c1})+1$ for $n_{\rm c1}\leq n_{\rm d} \lesssim n_{\rm c2}$. 
        }\vspace{\tlength}
\end{figure}
The transport measurements prove in the metallic impurity regime ($n_{\rm c1} <n_{\rm d} <n_{\rm c2}$) the delocalized nature of the electrons at the Fermi energy and as a consequence, $\tau_s$ is strongly influenced by the DP mechanism. However, towards lower densities $\tau_s$ is not any more calculated correctly from the transport data with a constant $\gamma_3$ (gray open triangles in Fig.~\ref{fig:dopdep}(a)) but only with a $\gamma_3$ that gradually changes with decreasing $n_{\rm d}$ from 6 to unity. Figure~\ref{fig:dopdep}(b) depicts the extracted values for $\gamma_3$ for which the calculated and measured $\tau_s^{-1}$ coincide for each doping density. For samples S6 and S5, whose doping densities are close to the MIT, the hopping induced spin relaxation (see below) is taken into account prior to the extraction of the DP related $\gamma_3$. The change of $\gamma_3$ is a direct consequence of the gradual change from small angle ionized impurity scattering to isotropic momentum scattering in the percolation regime with filamentary transport, i.e., the electron scattering changes with decreasing density from a Rutherford type scattering towards a thermal equilibrium Brownian motion with isotropic, random walk scattering. We want to point out in this context that the calculated density dependence of the total scattering rate given by the sum of Conwell-Weisskopf and percolation path scattering  perfectly matches the measured scattering rates (see Appendix).

At the MIT (samples S6 and S5), spin relaxation results from a mixture of delocalization and hopping of electrons between different impurity sites. The two contributions are both quantified by the transport experiments. In case of hopping induced spin relaxation the electron spin is rotated by a small angle $\theta (R_{ij} )$ during each hop due to spin orbit coupling quantified by $\gamma_{\rm D}$ \footnote{Here, we use the notation $\theta(R)$ instead of $\gamma(R)$ as in \cite{Dzh2002,Gor2003} in order to avoid confusion with the Dresselhaus constant $\gamma_{\rm D}$.}, i.e., motional narrowing does not apply in this case. The corresponding hopping induced spin relaxation is given by \cite{Dzh2002, Gor2003}:
\begin{equation}
\tau_\mathrm{s,hop}^{-1} =\tfrac{2}{3} \left\langle \theta^2(R_{ij} )\right\rangle/\tau_{\rm hop},
\label{eq:hop2}
\end{equation}
with the relation for $\langle \theta^2(R_{ij} )\rangle$ derived by \citet{Gor2003}.

For mainly localized electrons, the spin relaxation is dominated by the complex hyperfine interaction with the nuclei \cite{Lam1968, Ber2015}. This effect is diminished in the regime of variable range hopping by motional narrowing due to averaging of nuclear field configurations of different donor positions if the correlation time between hops is short in comparison to $\tau_s$. The hyperfine induced spin relaxation rate is given in this case by \cite{Kav2008}
\begin{equation}
\tau_{s,\mathrm{HF}}^{-1} = \left\langle \Omega_N^2\right\rangle\ \tau_{\rm c} = \left(\mu_\mathrm{B} g^*/\hbar \right)^2\left\langle B_N^2\right\rangle\,\tau_{\mathrm{c}},
\label{eq:HF}
\end{equation}
where $\left\langle B_N^2\right\rangle$ is the variance of the nuclear field fluctuation amounting for shallow donors in GaAs to $5.4\:$mT \cite{Dzh2002}. In the case of hopping, the correlation time $\tau_{\rm c}$ is clearly equal to the transport hopping time $\tau_{\rm hop}$, and again $\tau_s$ can be calculated without adjustable parameters. Taking into account hyperfine interaction and hopping processes, the spin relaxation times of the samples with $n_{\rm d} < n_{\rm c1}$ are reproduced extremely well using the afore determined values for $R_{\rm opt}$ and $\tau_{\rm hop}$.

\section{Conclusion}

In summary, Fig.~\ref{fig:dopdep}(a) shows the calculated (colored symbols) and the measured $\tau_s$ (black squares) which are for all doping densities in perfect agreement. We therefore conclude that the presented combination of detailed temperature dependent magneto-transport and spin dynamics measurements finally yields a complete \emph{quantitative} picture of the intrinsic electron spin relaxation around the metal-to-insulator transition. The experiments prove inter alia that the Dyakonov-Perel mechanism is not only valid above $n_{\rm c2}$ but also in the metallic impurity regime ($n_{\rm c1} < n_{\rm d} < n_{\rm c2}$)
where the scattering factor $\gamma_3$ changes with decreasing doping density linearly from small angle ionized impurity scattering to isotropic percolation path scattering.
In addition, the experimental data summarized in Fig.~\ref{fig:DopLit} show that the spin relaxation rates of the high quality samples yield a lower bound in respect to other data implying that the intrinsic spin relaxation rates can be strongly altered for example by surface depletion layers or by changes of the hopping rate due to charge compensation by background doping.
All the results are shown for GaAs but the basic principle of a twofold MIT with two critical doping concentrations ($n_{\rm c1}$ and $n_{\rm c2}$) is known in other semiconductors as well.\cite{Phi1998,Rin1995} While in III-V semiconductors the spin relaxation of delocalized electrons is governed by the DP mechanism presented here, group IV semiconductors, like silicon, are governed by the Elliot-Yafet mechanism instead.\cite{Cheng.PRL.2010} However, both processes depend on the electron scattering rate and are thus subject to the observed change in the scattering mechanisms presented here.

\begin{acknowledgments}
We acknowledge the financial support by the BMBF joint research project Q.com-H (16KIS0109 and 16KIS00107) and the Deutsche Forschungsgemeinschaft (OE 177/10-1) as well as from the NTH School for Contacts in Nanosystems.
\end{acknowledgments}

\appendix
\section{Sample and experimental data}

\begingroup
\squeezetable
\begin{table*}[tb]
\caption{Samples. The experimental doping densities $n_\mathrm{d}^{\rm exp}$ are obtained from high temperature Hall measurements. The mobilities and the corresponding diffusion constant are given for $6.5 \mathrm{K}$ by taking into account the two band model.}
\begin{tabular}{|l|c|c|c|c|c|c|c|c|c|c|}
\toprule
Sample No. & S1 & S2 & S3 & S4 & S5 & S6 & S7 & S8 & S9 & S10 \\
\toprule
$n_\mathrm{d}^{\rm exp}$ $\left(10^{16}\mathrm{cm^{-3}}\right)$ & 0.120(3) & 0.36(2) & 0.658(9) & 0.895(6) & 1.732(7) & 1.65(6) & 4.02(9) & 6.02(8) &8.20(5) & 10.31(5)\\
\colrule
$\mu_{\rm exp}^{6.5 \mathrm{K}}$ $\left(\mathrm{cm^{2}}\mathrm{(Vs)^{-1}}\right)$& --- & 66 (3) & 220 (20) & 370 (20) & 992 (2) & 944 (3) & 1870 (20) & 2330 (20) & 3160 (30) & 2653 (9) \\
\colrule
$D_{\rm exp}^{6.5 \mathrm{K}}$ $\left(\mathrm{cm^{2}}\mathrm{s^{-1}}\right)$& --- & 0.0368 (8) & 0.124 (3) & 0.208 (5) & 0.556 (1) & 0.529 (2) & 1.05 (1) & 1.305 (9) & 1.77 (2) & 1.486 (5) \\
\colrule
$A^0_\mathrm{II}$ $\left(10^{23} (\mathrm{mVs})^{-1}\right)$ & 1.1 (1) & 1.6 (2) & 1.27 (6) & 1.33 (6) & 1.08 (8) & 1.26 (3) & 1.32 (6) & 1.37 (7) & 1.43 (9) & 1.33 (2) \\
\colrule
$\mu^0_\mathrm{P}$ $\left(\rm m^2 (Vs)^{-1}\right)$ & 1.1 (1) & 0.75 (7) & 0.68 (2) & 0.77 (3) & 0.52 (2) & 0.505 (7) & 0.33 (2) & 0.34 (2) & 0.29 (2) & 0.32 (2) \\
\colrule
$E_E$ $\left(\mathrm{meV}\right)$ & 3.8 (2) & 3.3 (2) & 1.84 (8) & 1.06 (9) & 1.1 (5) & 1.0 (2) & 1.3 (2) & 1.1 (3) & 0.6 (6) & 0.0  (1) \\
\colrule
$\sigma_{m}^0$ $\left(1/\Omega \mathrm{cm}\right)$ & --- & --- & --- & --- & 1.47 (4) & 1.29 (3)& 10.0 (1) & 19.5 (2) & 42.8 (7) & 44.0 (2) \\
\colrule
$T_0$ $\left(10^2 K\right)$ & 826\footnotemark[1] & 143 (5) & 84 (2) & 46 (1) & 23.7 (8) & 23.3 (6) & --- & --- & --- & --- \\
\colrule
$\sigma_0$ $\left(1/\Omega \mathrm{cm}\right)$ & 2.4\footnotemark[1] & 5.8 (1) & 7.52 (6) & 10.1 (2) & 14.1 (3) & 14.2 (2) & --- & --- & --- & --- \\
\colrule
$\eta_\mu$ & 97(19) & 32(8) & 10.5(6) & 8.4(4) & 4.9(2) & 4.5(2) & --- & --- & --- & ---  \\
\colrule
$1/\tau_{\rm hop}^{6.5 \mathrm{K}}$ $\left(10^{11}/s\right)$ & 0.18\footnotemark[1] & 0.34 (2) & 1.44 (5) & 3.3 (2) & 93 (9) & 96 (7) & --- & --- & --- & --- \\
\colrule
$1/\tau_{\rm p}^{6.5 \mathrm{K}}$ $\left(10^{12}/s\right)$ & --- & --- & --- & --- & 36 (2) & 40 (2) & 14.0 (1) & 11.27 (8) & 8.30 (8) & 9.89 (3) \\
\botrule
\end{tabular}
\footnotetext[1]{Calculated using the extrapolated value of $N_\mathrm{E_F}$ from measurements on the higher doped samples.}
\label{tab:results}
\end{table*}
\endgroup

The MBE grown GaAs:Si epilayers are enclosed by an n-doped, 10~nm thick, top and bottom capping layer with a doping concentration of $n_{\rm d}=4\times 10^{18}$~cm$^{-3}$ and $n_{\rm d}=5\times 10^{16}$~cm$^{-3}$, respectively. The high doping concentration of the surface layer counteracts the effect of depletion due to Fermi level pinning at the surface. The bottom capping layer is separated from the substrate by a GaAs/AlGaAs superlattice and a 500~nm GaAs buffer layer. For the low doped samples (S1 to S5), an additional 500~nm thick Al$_{0.3}$Ga$_{0.3}$As barrier separates the top capping layer from the relevant n-doped epilayer to further reduce surface effects. All samples have been carefully designed by solving the one-dimensional Poisson equation self-consistently in order to obtain  the desired reduction of the depletion layer and to verify that the narrow cap layer has no effect on the transport measurements due to depletion by surface states. Table~\ref*{tab:results} summarizes the experimental doping densities $n_{\rm d}^{exp}$ extracted from the high temperature Hall measurements together with all other parameters obtained from transport.

\section{Measurement of the spin dynamics}

The spin relaxation times are measured by Hanle depolarization of the photoluminescence. Spin polarized electrons (and holes) are optically created in the sample by circular polarized light from a CW laser by above band gap excitation with a photon energy of $1.58$~eV. Application of a transverse magnetic field results in a Larmor precession of the continuously injected electron spins which yields an increasing randomization of the spin polarization $S$ and consequently a decreases of the PL polarization  with increasing transverse magnetic field $B$. The dependence $S(B)$ follows a Lorentz function where the half width at half maximum $B_{\sfrac{1}{2}}$ gives access to the spin relaxation time $\tau_\mathrm{s}$ \cite{Dyakonov.oo.1984} via the relation
\begin{equation*}
B_{\sfrac{1}{2}}=\frac{\hbar}{g^* \mu_{\rm B}} \left(\frac{1}{\tau_{l,e}}+\frac{1}{\tau_{s}}\right),
\end{equation*}
\begin{figure}[b]
    \includegraphics[width=\linewidth]{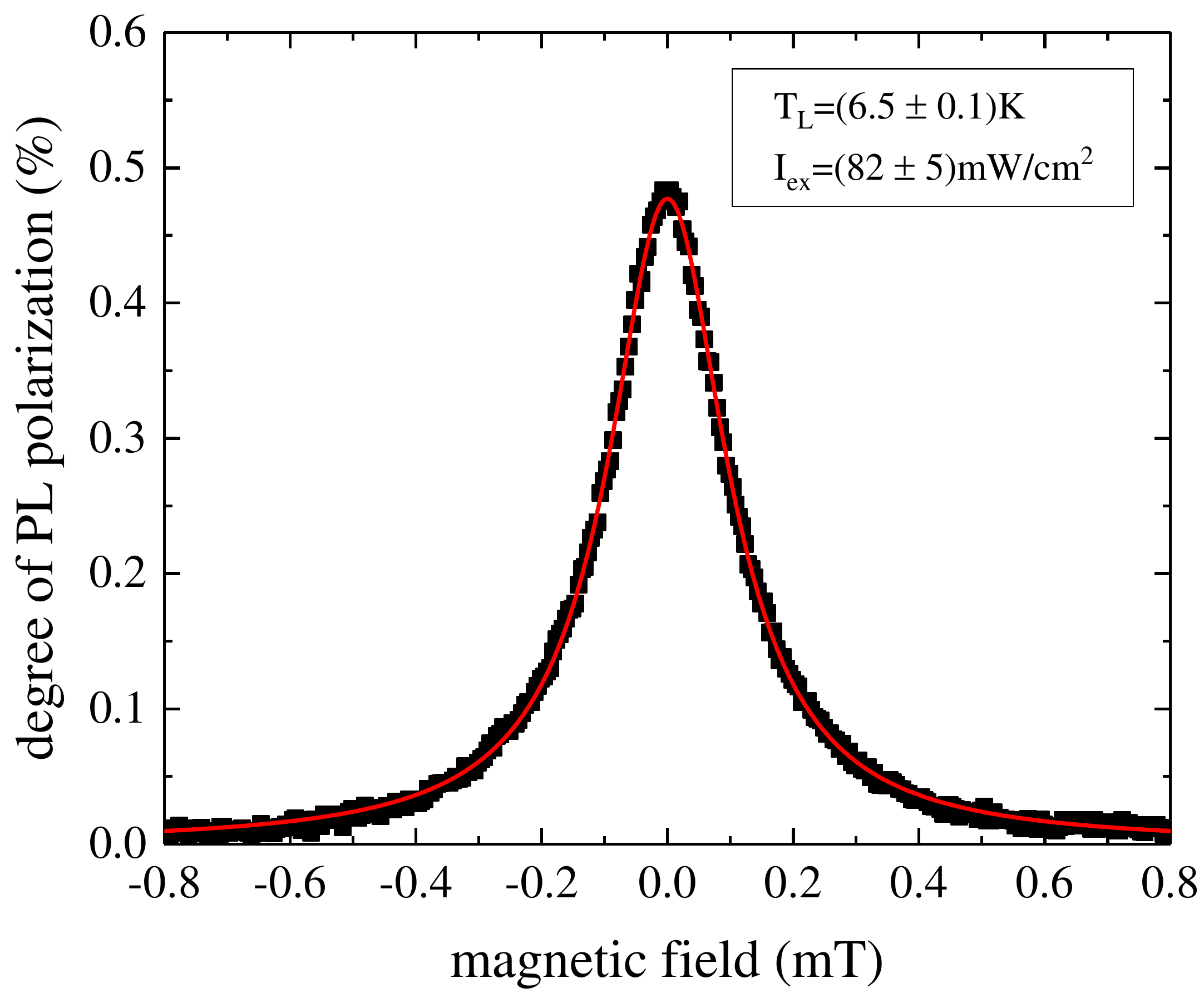}
        \caption{\label{fig:Hanle} (color online) A typical Hanle PL depolarization measurement. Applying a transverse magnetic field reduces the equilibrium spin polarization and thereby the degree of polarization of the PL. The red line is a Lorentzian fit to the data.}
\end{figure}
where $g^*$ is the effective electron g-factor and $\tau_{l,e}$ the effective radiative lifetime of the electrons. Non-radiative recombination does not play a role at low temperatures due to the high quality of the sample and the sample structure. Here,$\tau_{l,e}$ represents the mean time for the electron population to be fully replaced and strongly differs from the typical radiative lifetime which is governed here by the minority carriers, i.e., holes, in the present case for n-doped samples and low excitation densities. The energy dependence of $g^*$ is taken into account using the relation\cite{Hue2009} $g^*=-0.484+6.3\;eV^{-1}\times E$ . In general the Hanle depolarization technique is sensitive to the transverse spin dephasing time $\mathrm{T}_2$. However, for short correlation times $\tau_{\mathrm{c}}$ in the motional narrowing regime and at low magnetic fields spin relaxation and dephasing times are indistinguishable, i.e.,  $\mathrm{T}_1 = \mathrm{T}_2 = \tau_{s}$ (for a detailed discussion see, e.g., Ref. \cite{Zutic.RMP.2004,Heisterkamp.PRB.2015}). In all measurements the optically generated carrier density is kept about two orders of magnitude smaller than the intrinsic electron density in order to avoid any misleading impact of the photo-generated electrons on the intrinsic spin dynamics, i.e., $n_\mathrm{ex} \ll n_\mathrm{d}$. The radiative recombination maps the spin polarization of the conduction and impurity band electrons onto the polarization of the photoluminescence via the optical selection rules. The hole spins undergo a much faster spin relaxation compared to the electrons due the strong spin-orbit interaction \cite{Dzh2002} such that for all measurements the hole spins are in very good approximation unpolarized. Hence, in the low excitation regime the width of the Hanle curve is dominated by the electron spin relaxation time $\tau_{s}$ which is much shorter than $\tau_{\rm l, e}$. In order to further diminish any remaining influence of our excitation, the measured values of $B_{\text{\sfrac{1}{2}}}$ are extrapolated towards zero excitation density in order to obtain the intrinsic electron spin relaxation time $\tau_{s}$.

An effective nuclear spin polarization due to optical pumping can significantly influence the electron spin dynamics \cite{Dya2006}. To this end, the exciting laser light is modulated between $\sigma^+$ and $\sigma^-$ polarization using an optical modulator in order to prevent the build-up of any nuclear polarization. From modulation frequency dependent measurements $\rm \left(2\,Hz - 500\,kHz\right)$ (not shown) we conclude that there is no significant nuclear polarization above $\rm 2\,\mathrm{kHz}$. All presented measurements were performed at $\rm 50kHz$.

\section{Momentum scattering in the impurity band}

\begin{figure}[b]
	\centering
		\includegraphics[width=\columnwidth]{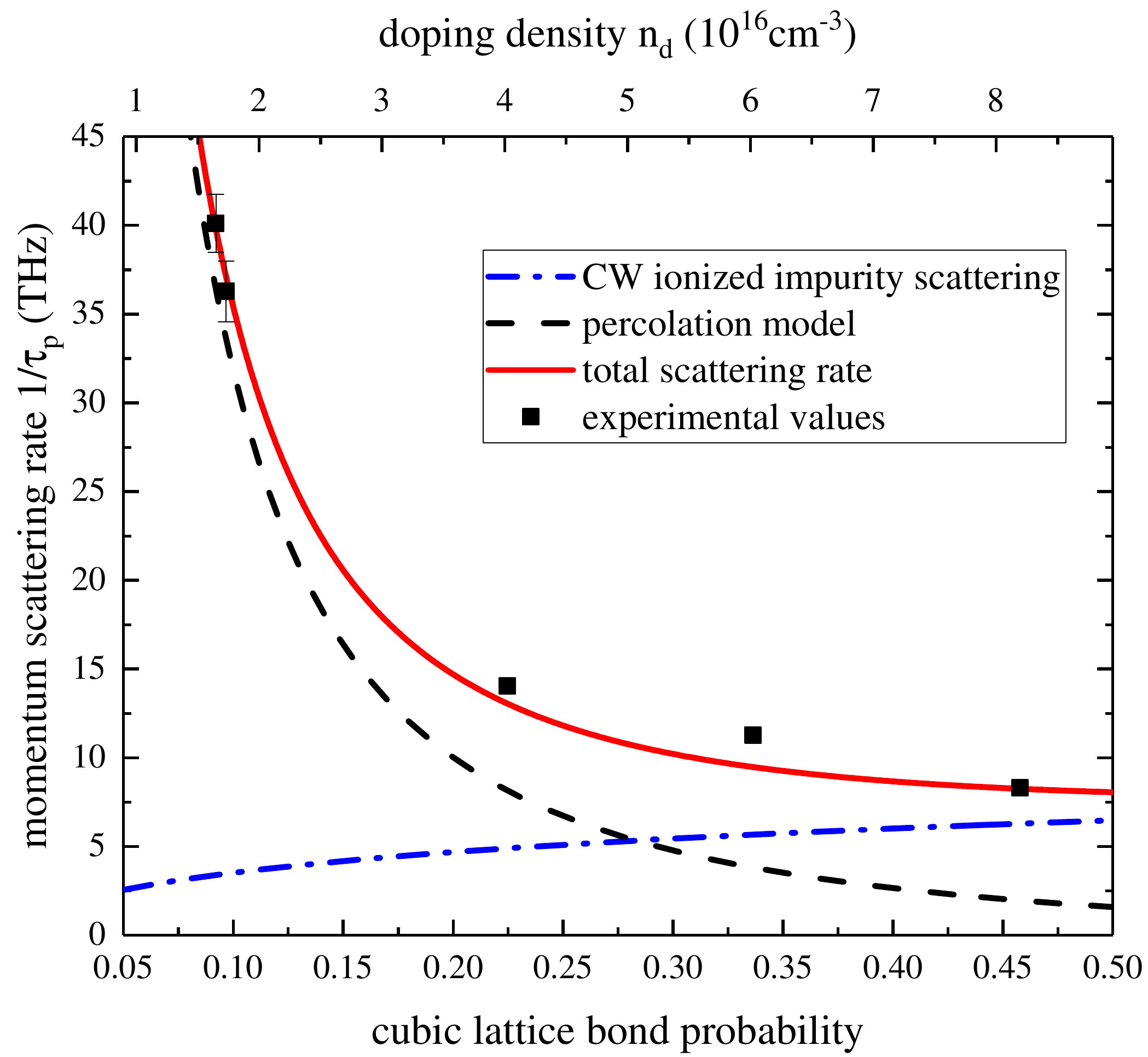}
	\caption{(color online) Impurity band scattering rate as a function of the probability of finding another donor on a reduced cubic percolation lattice.}
	\label{fig:IBpercolation}
\end{figure}

The momentum scattering of electrons in the impurity band results for densities between $n_{\rm c1}$ and $n_{\rm c2}$ from ionized impurity scattering and percolation path scattering. \cite{Moo2009} The ionized impurity scattering is in this regime well described by the Conwell-Weiskopf approach. Percolation path scattering is in general much more complex and requires quantum mechanical, tight-binding bond percolation models which can be solved numerically only. However, \citet{Sch2014} have shown recently that the full quantum mechanical calculations of conductivity and mean free path are in good accord with results from very simple heuristic considerations. They successfully describe for a three dimensional cubic percolation lattice the mean free path $\lambda$ by
\begin{equation}
\lambda\propto\frac{p^2}{1-p},
\label{eq:meanfreepath}
\end{equation}
where $6\cdot p$ is the mean number of connected neighbors, i.e., the bond probability. \cite{Sta2009} We estimate $p$ for our specific case by calculating the average number of donors found in a sphere of radius $2 a_\mathrm{d}$ and divide by the 6 possible translation directions:
\begin{equation}
p=\frac{1}{6}\left(4/3 \pi \left(2a_{\rm d}\right)^3\right)n_{\rm d},
\label{eq:prop}
\end{equation}
where $a_{\rm d}$ denotes the Bohr radius of the donor-bound electron. \citet{Sch2014} also estimate from their quantum mechanical model that the square of the mean particle velocity follows $v^2\propto p$. Thus, we obtain for the momentum scattering rate due to percolation
\begin{equation}
\tau_{p}^{-1} =v/\lambda \propto \frac{1-p}{p^{3/2}}.
\label{eq:scatteringtime}
\end{equation}
Next, we compare the calculated with our experimental momentum scattering rates whereby the proportionality factor in Eq.~\ref{eq:scatteringtime} is the only fitting parameter. Figure~\ref{fig:IBpercolation} depicts $1/\tau_p$ for CW ionized impurity scattering (dashed dotted line), percolation path scattering (dashed line), and experiment (black squares). The solid red line in Fig.~\ref{fig:IBpercolation} depicts the calculated total momentum scattering rate which is in excellent agreement with the experiment, i.e., the functional relation of the percolation path scattering is well confirmed by the experiment. The results also prove that ionized impurity scattering dominates close to $n_{c2}$ while percolation path scattering dominates in the regime of $n_{c1}$.

\end{document}